# Leveraging Anatomical Constraints with Uncertainty for Pneumothorax Segmentation

Han Yuan[1†], Chuan Hong[2†], Nguyen Tuan Anh Tran[3], Xinxing Xu[4], Nan Liu[1,5,6✉]

[1] Centre for Quantitative Medicine, Duke-NUS Medical School, Singapore
[2] Department of Biostatistics and Bioinformatics, Duke University, USA
[3] Department of Diagnostic Radiology, Singapore General Hospital, Singapore
[4] Institute of High Performance Computing, Agency for Science, Technology and Research, Singapore
[5] Programme in Health Services and Systems Research, Duke-NUS Medical School, Singapore
[6] Institute of Data Science, National University of Singapore, Singapore

[†] Equal contribution

[✉] Correspondence: Nan Liu, Centre for Quantitative Medicine, Duke-NUS Medical School, 8 College Road, Singapore 169857, Singapore
Phone: +65 6601 6503
Email: liu.nan@duke-nus.edu.sg

**Abstract**

Background
Pneumothorax is a medical emergency caused by abnormal accumulation of air in the pleural space – the potential space between the lungs and chest wall. On 2D chest radiographs, pneumothorax occurs within the thoracic cavity and outside of the mediastinum and we refer to this area as "lung+ space". While deep learning (DL) has increasingly been utilized to segment pneumothorax lesions in chest radiographs, many existing DL models employ an end-to-end approach. These models directly map chest radiographs to clinician-annotated lesion areas, often neglecting the vital domain knowledge that pneumothorax is inherently location-sensitive.

Methods
We propose a novel approach that incorporates the lung+ space as a constraint during DL model training for pneumothorax segmentation on 2D chest radiographs. To circumvent the need for additional annotations and to prevent potential label leakage on the target task, our method utilizes external datasets and an auxiliary task of lung segmentation. This approach generates a specific constraint of lung+ space for each chest radiograph. Furthermore, we have incorporated a discriminator to eliminate unreliable constraints caused by the domain shift between the auxiliary and target datasets.

Results
Our results demonstrated significant improvements, with average performance gains of 4.6%, 3.6%, and 3.3% regarding Intersection over Union (IoU), Dice Similarity Coefficient (DSC), and Hausdorff Distance (HD). These results were consistent across six baseline models built on three architectures (U-Net, LinkNet, or PSPNet) and two backbones (VGG-11 or MobileOne-S0). We further conducted an ablation study to evaluate the contribution of each component in the proposed method and undertook several robustness studies on hyperparameter selection to validate the stability of our method.

Conclusion
The integration of domain knowledge in DL models for medical applications has often been underemphasized. Our research underscores the significance of incorporating medical domain knowledge about the location-specific nature of pneumothorax to enhance DL-based lesion segmentation. Different from previous work that requires additional annotation on the target dataset and task, our method leverages external publicly available datasets and an auxiliary task.

This adaptation makes our approach a promising strategy for diagnosing other thoracic conditions that possesses location-relevant characteristics.

# 1. Introduction
*1.1 Motivation*

Pneumothorax is a medical emergency caused by the abnormal accumulation of air in the pleural space, which is the potential space between the lungs and chest wall [1-3]. The pleural space is non-existent in healthy controls, whereas it can occupy significant portions of the thoracic cavity in patients with a large pneumothorax. Chest radiographs serve as the primary diagnostic tool for pneumothorax, aiding in identifying its location and estimating its size [2, 4]. Traditionally, radiologists report chest radiographs based on their domain knowledge and past experience [5-7]. In recent years, with the advent of deep learning (DL), there has been a shift towards automated detection and segmentation of pneumothorax from chest radiographs, achieving promising performance when paired with high-quality annotations [8, 9]. Beyond image-level classification [10], DL offers fine-grained segmentation of lesion, further assisting non-radiologist clinicians lacking systematic radiological training to make initial diagnoses [11]. However, many existing DL models employ an end-to-end approach [12], often neglecting the fact that pneumothorax is inherently location-sensitive, predominantly manifesting within the pleural space [10, 11, 13-17]. Our approach aims to leverage the medical domain knowledge of disease occurrence into DL-based pneumothorax segmentation, emphasizing the significance of the location-specific nature of the disease.

*1.2 Related work*
*1.2.1 Computer-aided diagnosis for pneumothorax*

Computer-aided tools have emerged as promising solutions for diagnosing pneumothorax from chest radiograph. Numerous studies have reported accurate classification using various DL models [18], such as AlexNet [19, 20], VGG [19, 21], Inception [19, 22], EfficientNet [23-25], Network in Network [26], ResNet [24, 27], DenseNet [19, 28], AlbuNet [13, 29], Spatial Transformer [10, 30], and others. While most studies employ the conventional end-to-end approach, Chen et al. designed a two-stage model for pneumothorax classification [31], combining an object detection model YOLO [32] and classification techniques ResNet [27] and DenseNet [28] to distinguish pneumothorax patients and healthy individuals based on the cropped lung field.

Recent research has expanded beyond merely image-level labeling, emphasizing pixel-level lesion area delineation. This not only reduces radiologist' workload but also fosters greater trust in automated systems [18, 33, 34]. Like the binary classification of pneumothorax, DL-based techniques are the state-of-the-art choices for pneumothorax area segmentation. The techniques used mainly comprise the architectures of U-Net [11, 15, 16, 35], DeepLab [11, 36], Mask R-CNN [11, 37], with the backbones of ResNet [11, 15, 16], SE-ResNext [11, 16], DenseNet [14, 16], EfficientNet [11]. Apart from tailoring models on specific datasets, commercial systems, such as DEEP: CHEST-XR-03, have also been developed and clinically validated for pneumothorax segmentation [38].

*1.2.2 Domain knowledge in chest radiograph analysis*

While most of the models mentioned above employ an end-to-end approach, mapping input radiographs directly to target disease labels or lesion annotations, the integration of domain knowledge in medical image analysis is pivotal. Incorporating such knowledge not only enhances model interpretability but also boosts model performance [39, 40]. In chest radiograph analysis, disease localization serves as invaluable prior knowledge. For example, Li et al. identified specific anatomical regions within the lung zone, highlighting their diagnostic relevance in detecting conditions cardiomegaly and pleural effusion [41]. Similarly, Crosby et al. focused on the upper third of chest radiograph and demonstrated that a VGG-based classifier trained barely on the sub-region yielded a good performance in the task of distinguishing pneumothorax [42]. Recently, Jung et al. proposed to extract the domain knowledge of thoracic diseases occurrence from the class activation map of a thoracic disease classifier. The extracted knowledge was then used for precise thoracic disease localization [43]. Such methods typically utilize masks at various stages, from input to output, ensuring the model's attention is directed towards areas with a higher likelihood of disease presence. More recently, Bateson et al. introduced an innovative method that incorporates prior knowledge about the organ size into the model training [39, 44]. By penalizing deviations from this domain knowledge, the model is guided to align more closely with clinical insights. Notably, in tasks of spine and heart segmentation, their method achieved considerably better performance than conventional methods.

*1.3 Contribution*

According to the clinical knowledge [1-3], pneumothorax is an abnormal gas collection in the pleural space, which is a potential space between the lung and chest wall. On a 2D projection of a chest radiograph, pneumothorax localizes inside the thoracic cavity and outside of the mediastinum. We refer to this area as "lung+ space", which includes the lung and the pleural space on a chest radiograph. We propose to incorporate the location information into pneumothorax segmentation. Inspired by [39, 44, 45], we employed occurrence information as a guiding constraint in the segmentation training. To circumvent the need for additional domain knowledge annotation, we proposed a three-phase pipeline to determine the disease occurrence area. Leveraging both external open-access datasets and existing annotations, our evaluations, conducted on a widely-recognized pneumothorax segmentation dataset, revealed that our constrained training approach consistently outperformed traditional methods across a range of architectures and backbones. Moreover, to validate the efficacy of our approach, we have conducted an ablation study and two robustness experiments. These studies underscored the effectiveness of individual components within our design and affirmed the overall stability of our proposed method. We hope the proposed method provides a useful framework for embedding domain knowledge into the diagnosis of other thoracic conditions that possesses location-relevant characteristics.

**2. Method**

To harness the power of domain knowledge in improving DL efficacy in lesion segmentation, we propose a three-phase pipeline that incorporates disease occurrence knowledge into the training stage of the pneumothorax segmenter. In this section, we first provide an overview of the proposed pipeline. We then detail the generation process of anatomical constraints. Finally, we elaborate the constrained training approaches for lesion segmentation.

*2.1 Overview of the proposed pipeline*

As outlined in **Figure 1**, the proposed pipeline contains three indispensable phases to obtain sample-specific lung+ space, select well-behaved lung+ space, and implement constrained training of pneumothorax segmenter. In Phase 1, we develop an auxiliary lung segmenter using three public lung segmentation datasets, including Japanese Society of Radiological Technology dataset (JSRT) [46], Shenzhen dataset (Shenzhen) [47], and Montgomery County dataset (MC) [47]. This segmenter is then integrated with morphological operations, including connected component cutoff, closing, and dilation to derive a lung+ space segmenter. This refined segmenter is subsequently deployed on the target dataset of pneumothorax segmentation to predict lung+ space. In Phase 2, we introduce a lung+ space discriminator, crafted using the training and validation dataset for the pneumothorax segmentation, filtering out inaccurately predicted lung+ spaces from Phase 1, ensuring only high-quality constraints are retained. In Phase 3, with the selected lung+ space from Phase 2, we proceed to train the pneumothorax segmenter using a constrained approach. To detail the three phases, the remaining parts are organized as follows: First, we provide in **Section 2.2** the design details of Phases 1 and 2 for generating the well-behaved anatomical constraints for pneumothorax, including three coherent modules of lung area segmenter, lung+ space segmenter, and lung+ space discriminator. Second, we illustrate the formulation of a standard lesion segmenter and outline the proposed constrained segmenter of Phase 3 in **Section 2.3**.

*2.2 Learning the anatomical constraints*

The primary challenge in our proposed method lies in the derivation of well-behaved constraints. From the domain knowledge, we understand that pneumothorax occurs in the lung+ space and the lung+ space can be used as a constraint in segmenter training. The previous work generally requires constraint-relevant annotation on the target dataset, which incurs additional expenses and lacks adaptability when extended to other datasets [39, 44, 45, 48, 49]. To tackle this problem, we exploit three external open-access datasets and an auxiliary task of lung segmentation to generate the lung+ space as our anatomical constraints. Specifically, we first generate a lung area segmenter based on the external datasets. Then, we integrate several post-processing techniques into the lung area segmenter to achieve a lung+ space segmenter and deploy it on the target dataset. Finally, considering the inherent noise in the constraints generated from external source-based model, a lung+ space discriminator is developed to filter out the noisy constraints while retaining the informative ones for the constrained training of the pneumothorax segmenter.

*2.2.1 Phase 1: Lung area segmenter*

The foundation for our anatomical constraints learning is the lung area segmenter trained on external datasets. We utilize three publicly available chest radiograph datasets with lung area annotations and developed the lung area segmenter following the standard area segmentation [46, 47]. An intuitive idea is to transfer the trained lung area segmenter to the target dataset, obtain the lung area, and use them as the constraints. However, according to domain knowledge, pneumothorax is abnormal gas (lucent area) in the pleural space, which is localized to the lung+ space on 2D chest radiographs. Additionally, the directly deployed lung segmenter suffers from domain shift between the source dataset and the target dataset [39, 48, 50]. Consequently, certain portions of the generated lung area may exhibit detrimental noise to the downstream pneumothorax segmentation. To tackle the first challenge, we integrate several morphological operations into the fundamental lung area segmenter to derive the lung+ space segmenter. To address the second issue, we design a lung+ space discriminator to filter out the noisy constraints of lung+ space and retain the informative ones.

*2.2.2 Phase 1: Lung+ space segmenter*

Based on the auxiliary lung segmenter, we implement three post-processing steps, including the largest connected component cutoff, morphological closing, and morphological dilation to develop a lung+ space segmenter. The initial operation, the largest connected component cutoff, is employed to mitigate the impact of small islands, a common issue in medical image segmentation [51, 52]. By retaining the top two largest areas, we effectively extract the two frontal lung regions from the noisy segmentation output. Subsequently, morphological closing is applied to fill small holes inside the lung area, another prevalent concern in medical image segmentation [51, 53, 54]. Lastly, morphological dilation is employed to extend the lung area, encompassing the side pleural space between the lung boundary and the chest wall [55]. By combining these post-processing operations with the lung area segmenter, we have effectively crafted the lung+ space segmenter which generates the candidate constraints of lung+ space for each radiograph in the target dataset of pneumothorax segmentation.

*2.2.3 Phase 2: Lung+ space discriminator*

Utilizing the lung+ space segmenter, each radiograph $I_i$ in the target task has been enriched with a constraint $C_i$ Which holds an identical shape as $I_i$. In a constraint $C_i$, pixels located within the lung+ space are assigned a value of 1, while those outside the lung+ space are marked as 0. Nevertheless, due to the problem of domain shift [39, 48], some of the constraints still present noise issues. If all constraints are unquestioningly incorporated in model training, the improvements achieved by well-behaved constraints can potentially be negated by the detrimental constraints. To address this challenge, a lung+ space discriminator is developed to differentiate whether a predicted lung+ space contains significant noise.

The constraint discrimination constitutes a binary classification task. For training the discriminator, binary labels are derived by calculating the coverage rate $R_i$ of pneumothorax annotation $S_i$ in relation to the lung+ space constraint $C_i$. If $R_i$ exceeds $\tau$, the binary label $B_i$ is designated as 1; otherwise, it is set to 0.

$$R_i = \frac{|C_i \cap S_i|}{|S_i|} \tag{1}$$

Regarding the discriminator input, we followed a previous study [56] to fuse the original $I_i$, the constraint $C_i$, and the masked image $I_i \times C_i$ on the channel-level. The model output is configured to yield a single value, predicting the probability of including the constraint in the training phase. With the inputs, output targets, and the binary classifier, we proceed with the standard image classification using cross-entropy loss [55]. After the binary classification, certain radiographs are augmented with the constraints of lung+ space. For those samples whose constraints are excluded, we supplement them with all-ones matrixes to align data format and nullify constraint effect. Finally, we obtain the anatomical constraints of lung+ space for the downstream training of the constrained segmenter. It is noteworthy that the classification is performed exclusively on the training and validation sets of the target datasets, ensuring that the test set is unseen and there is no information leakage.

*2.3 Constrained medical image segmentation*

We first present a standard formulation of image segmenter training without any constraints, which serves as the baseline for our study. Subsequently, we elaborate on the constrained training approach, which includes an additional penalty term aimed at satisfying the imposed constraints.

In a typical training stage of a single disease segmenter, we consider a dataset $D$ consisting of $N$ input images $I_i, i = 1, \ldots, N$ and their respective lesion segmentations $S_i, i = 1, \ldots, N$. Then the dataset $D$ is randomly split into training set $D_{train}$, validation set $D_{valid}$, and test set $D_{test}$ with $N_{train}$, $N_{valid}$, and $N_{test}$ samples, respectively. After that, a model $Y$ with parameter $\theta$ is trained on $D_{train}$, by minimizing the overall loss $L$ averaging sample-wise loss $l$ such as Dice or cross-entropy [57] between the model output $Y(I_i, \theta)$ and the ground-truth mask $S_i$.

$$L_{D_{train}} = \frac{1}{N_{train}} \sum_{i=1}^{N_{train}} l(Y(I_i, \theta), S_i) \tag{2}$$

To avoid the overfitting of $Y$, the optimization of $\theta$ is early-stopped upon reaching the loss plateau on $D_{valid}$. However, this approach overlooks the domain knowledge and trains the lesion segmenter through an end-to-end approach to map $I_i$ to $S_i$. In clinical medicine, some diseases highly occur in certain regions, which receive more attention than other area in the diagnostic process [55]. Therefore, the prior knowledge of disease occurrence area potentially contributes to the training process of disease segmenter.

*2.3.1 Phase 3: Constrained lesion area segmenter*

In our proposed formulation, the disease occurrence area is introduced as a constraint in loss function. To guide the model focus on the disease occurrence area, the adjusted loss function penalizes model if the constraint is violated. Specifically, the loss function in **Formula (2)** will be supplemented with a novel penalty term $P$ comparing the model output $Y(I_i, \theta)$ with a sample-specific constraint $C_i$:

$$L_{D_{train}} = \frac{1}{N_{train}} \sum_{i=1}^{N_{train}} (l(Y(I_i, \theta), S_i) + \lambda \times P(Y(I_i, \theta), C_i)) \tag{3}$$

where $l(Y(I_i, \theta), S_i)$ stands for the classic loss function for image segmentation in **Formula (2)**, $\lambda$ is a positive hyper-parameter fine-tuned on $D_{valid}$, and $P(Y(I_i, \theta), C_i)$ denotes the proposed penalty term which takes the following form:

$$P(Y(I_i, \theta), C_i) = 1 - \frac{|Y(I_i, \theta) \cap C_i|}{|Y(I_i, \theta)|} \tag{4}$$

with $|Y(I_i, \theta) \cap C_i|$ standing for the intersection area size between $Y(I_i, \theta)$ and $C_i$, and $|Y(I_i, \theta)|$ representing the size of $Y(I_i, \theta)$. Clearly, when an outputted disease segmentation crosses the constraint boundary, the penalty function is positive while a satisfied segmentation within the constraint area corresponds to a null penalty. Through adding such a differentiable term in loss function, we address the difficulty of conventional convex-optimization of neural networks for the constraint satisfaction [39, 58, 59].

# 3. Experiments

## 3.1 Experimental settings

### 3.1.1 Datasets

Our proposed formulation was based on three external datasets of lung segmentation and a target dataset of pneumothorax segmentation. **Table 1** gives an overview of the used datasets and their purposes in our experiments. The lung segmentation datasets were used to develop the auxiliary lung segmenter, foundation for the lung+ space segmenter. The pneumothorax dataset was our target dataset to build the lung+ space discriminator and compare the pneumothorax segmenter trained by the baseline or the constrained loss function.

*Lung segmentation dataset*: We developed the lung segmenter using Japanese Society of Radiological Technology dataset (JSRT) [46], Shenzhen dataset (Shenzhen) [47], and Montgomery County dataset (MC) [47]. The JSRT dataset was collected by the Japan Radiological Society and it includes 247 chest radiographs with a resolution of 2048×2048. The original dataset contained no ground-truth lung area annotation, which was supplemented by another team [60]. The MC dataset was gathered by Montgomery County's tuberculosis screening program and it comprises of 138 chest radiographs with either a resolution of 4020×4892 or 4892×4020. The Shenzhen dataset was built up by Shenzhen No. 3 Hospital and it consists of 566 chest radiographs with varying resolutions around 3000×3000. Both MC and Shenzhen dataset were annotated with lung area masks by researchers from U.S. National Institutes of Health [47]. We resized all radiographs from the three datasets and their corresponding lung area annotations into the resolution of 224×224 to comply with most pre-trained backbones [21, 54, 61]. Then we employed the proportion of 70/20/10 for the split of training (173 JSRT radiographs, 396 Shenzhen radiographs, 97 MC radiographs), validation (49 JSRT radiographs, 113 Shenzhen radiographs, 28 MC radiographs), and test dataset (25 JSRT radiographs, 57 Shenzhen radiographs, 13 MC radiographs). We included more radiographs in the validation dataset than the test dataset per the auxiliary task aim for training a good foundation model.

*Pneumothorax segmentation dataset*: With the auxiliary lung segmenter built on the external datasets, we trained the lung+ space discriminator, the baseline pneumothorax segmenter, and the constrained pneumothorax segmenter on the Society for Imaging Informatics in Medicine (SIIM)-American College of Radiology (ACR) Pneumothorax Segmentation dataset (SIIM-ACR) [62]. As a subset the ChestX-ray14 dataset [63], the SIIM-ACR dataset was prepared by radiologists from SIIM and Society of Thoracic Radiology (STR). It comprises 2391 pneumothorax-positive radiographs and the matching lesion annotations with a resolution of 1024×1024. All radiographs and lesion annotations were resized into 224×224 and further split into training (1674 radiographs), validation (239 radiographs), and test dataset (478 radiographs) using the ratio of 70/10/20 to involve more samples in the test set and ensure the robustness of evaluation.

*3.1.2 Baseline segmentation*

To demonstrate the impact of constraints, we employed six segmentation networks using three architecture of U-Net [35], LinkNet [64], or PSPNet [65], along with two backbones of VGG-11 [21] or MobileOne-S0 [61] to serve as the baseline. The model input was the grayscale chest radiograph, replicated three times on the channel-level to align with most backbones. The model output was a standard single-channel probability map with a resolution of 224×224, which was binarized using a common threshold of 0.5 to delineate the specific disease region [66]. The standard Dice loss was used in model training [57].

*3.1.3 Anatomical constraints learning*

To learn well-behaved constraints of lung+ space without additional annotation in the target dataset, we followed a multi-step approach: First, we extracted constraints from an auxiliary task involving lung area segmentation using three external datasets; Next, we applied three morphological operations to transform the lung area into the lung+ space; Finally, we trained a lung+ space discriminator to filter out noisy constraints.

The lung segmenter in the first step was constructed based on the U-Net architecture with the VGG-11 backbone. The three morphological operations in the second step consisted of the following: the top two largest connected component cutoffs, closing with a 19×19 ellipse element [51] to fill voids within the lung area, and dilation with a 15×15 ellipse element [51] to smooth the lung area boundary and encompass the lung+ space outside the lung area. The lung+ space discriminator in the third step was based on the VGG-11 backbone, and we adjusted its output to produce a single predictive value. The coverage rate $\tau$ was set as 0.99 in the generation of classification label $B_i$. A further step for noise filtering was to increase the binarization cutoff value to include the constraints with high-confidence. We chose the cutoff value based on several high specificity values of 0.80, 0.85, 0.90, and 0.95 and subsequently optimized it as a hyper-parameter in the constrained segmentation training.

*3.1.4 Constrained segmentation*

After the selection via the lung+ space discriminator, the well-behaved constraints were kept and for those samples without constraints, an all-ones matrix without penalty effect was used to ensure the data alignment in training process. We modified the loss function in model training by adding a penalty term to the standard Dice loss. The penalty term compared each pneumothorax prediction with its corresponding constraint. Hyper-parameter $\lambda$ in the modified loss function was grid selected from 0.2, 0.4, 0.6, 0.8, and 1.0 according to model performance on the validation set. Despite the modification of loss function, all other settings remained unchanged as the baseline models, including the model inputs, outputs, architectures, backbones, and optimizer, to facilitate a fair comparison.

*3.1.5 Implementation details*

We utilized the standard Dice loss for training the lung area segmenter and the baseline pneumothorax segmenter. For the training of the lung+ space discriminator, we employed the classical cross-entropy loss. For the constrained training of pneumothorax segmenter, we added a penalty term to the standard Dice loss. Across all experiments, Stochastic Gradient Descent (SGD) was employed to minimize the respective loss functions. We initiated the learning rate at 0.01 and reduced it to 0.9 of its current value if no improvement was observed for 5 epochs on the validation dataset. **Table 2** provides an overview of the experimental details.

For reproducibility, the pipeline was developed in PyTorch 1.12.1 and the code has been open access [67]. We implemented the experiments on a Dell Precision 7920 Tower Workstation with an Intel Xeon Silver 4210 CPU and an NVIDIA GeForce RTX 2080 Super GPU.

*3.1.6 Evaluation metrics*

*Evaluation on constraints plausibility*: The anatomical constraints of lung+ space, generated by the external lung segmenter and morphological operations, contained high uncertainty. To filter out the noisy constraints, we trained a lung+ space discriminator and used Area Under the Receiver Operating characteristic Curve (AUROC) to assess the model performance. We set the classification thresholds according to the specificity values of 0.80, 0.85, 0.90, 0.95 [68] and reported the sensitivity, positive predictive value (PPV), and negative predictive value (NPV) to present the characteristics of predicted constraints after selection. Beside the evaluation metrics, we also reported their respective standard errors via bootstrapping.

*Evaluation on segmentation performance*: We utilized the Intersection over Union (IoU) [69], Dice Similarity Coefficient (DSC) [70], and the Hausdorff Distance (HD) [71] to assess the pixel-level difference between the predicted area and the ground-truth annotation in both the auxiliary task of lung area segmentation and the target task of pneumothorax segmentation. IoU and DSC are designed to quantify the degree of overlap between the predicted area and the ground-truth annotation. Accordingly, elevated values in these two metrics indicate enhanced model performance. On the other hand, HD assesses the distance between the two aforementioned regions, and therefore, diminished HD stands for better performance. In addition to the mean values on the test dataset, we also provided the standard errors based on bootstrapping.

*3.2 Results*

This section presents quantitative and qualitative results of each module in the proposed constrained segmentation. For the quantitative evaluations, we reported the results of auxiliary lung segmentation, lung+ space discrimination, and pneumothorax segmentation. An ablation study was presented to underscore the efficacy of each designed element. We further highlighted the robustness study to validate the stability of our constraint-based formulation. For the qualitative assessments, we first showed the visual samples of lung segmentation derived from the external lung segmentation dataset. We then presented the various phases involved in lung+ space generation. Lastly, we compared the pneumothorax segmenter trained by using baseline and that trained using our constrained loss function.

*3.2.1 Auxiliary lung area segmentation*

**Table 3** shows the auxiliary lung segmentation results using the U-Net architecture and the VGG-11 backbones. The segmenter yielded best IoU, DSC, and HD on the test dataset of MC database and performed comparably good on the other two databases. **Figure 2** visualizes three random samples and their segmentation results, demonstrating the outstanding performance of the auxiliary lung area segmenter on the external datasets.

*3.2.2 Lung+ space discriminator*

To obtain the lung+ space, we first applied the auxiliary lung segmenter on the target dataset of pneumothorax segmentation to obtain the initial lung area and then processed the lung area with three morphological operations. The first operation selected the top two largest connected components, thereby eliminating noisy segment islands [72]. The second operation entailed a closing operation using an element size of 19×19 to fill voids within the lung area. The final operation consisted of dilation with 15×15 element size to encompass the pleural space situated between the lung and chest wall.

**Figure 3** presents different stages of lung segmentation and morphological operations on the SIIM-ACR dataset. Radiographs in the first two role demonstrates the noisy small islands and the holes in lung structure, which is alleviated by the post-processing methods. However, it is worth noting that not all constraints can be adjusted, as depicted in the last two rows, which showcase collapsed scenarios and validate the essential role of the reliability discriminator in identifying and discarding erroneous constraints.

Based on the predicted lung+ space, we compared it to the ground-truth pneumothorax annotation. A lung+ space was labeled well-behaved if it covered 0.99 of its corresponding pneumothorax areas. With these binary labels, a discriminator was trained and tested. To further mitigate the introduction of noisy constraints in downstream training, we established cutoff values based on multiple specificity thresholds. **Table 4** presents the discriminator's performance in terms of AUROC, specificity-based cutoff values, the respective specificity, sensitivity, PPV, and NPV values.

*3.2.3 Pneumothorax segmentation*

**Table 5** quantitatively compares the constrained and the baseline segmentation performance across different combinations of architectures and backbones. The constrained version consistently outperformed the baseline method and yielded average performance gains of 4.6%, 3.6%, and 3.3% in terms of IoU, DSC, and HD. Across all baseline models, U-Net architecture with VGG-11 backbone achieved best performance in terms of IoU and DSC. Adopting the constrained training strategy, the same combination obtained the state-of-the-art results of IoU, DSC, and HD across all baseline and constrained models. Last, U-Net architecture achieved better results compared with the relatively sophisticated LinkNet or PSPNet in most scenarios, demonstrating the effectiveness of U-Net for pneumothorax segmentation.

**Figure 4** provides a comparative visualization of pneumothorax segmentation between the constrained and baseline segmenters using the architecture of U-Net and the backbone of VGG-11. The first row exemplifies a scenario where the proposed method surpasses the baseline model, the second row presents a scenario where the proposed method achieves performance on par with the baseline model, and the third row outlines a situation where the proposed method underperformed the baseline model. The last row illustrates cases where both methods encounter difficulties.

*3.2.4 Ablation study*

To ascertain the efficacy of each component within the proposed formulation, we conducted an ablation study on the U-Net architecture and the VGG-11 backbone. As shown in **Table 6**, the constraints generated by the external lung area segmenter exhibited excessive noise, resulting in subpar performance compared to the baseline without any constraints. While training with constraints derived from the lung+ space segmenter yielded better performance, it only achieved performance marginally better than the baseline. In contrast, with the integration of a lung+ space discriminator, the noisy constraints were efficiently filtered, preserving the well-behaved ones, and leading to substantial improvements across all evaluation metrics.

*3.2.5 Robustness to the constraint settings*

Finally, we investigated the robustness of hyperparameters in constraints learning. To this end, we validated a range of values in the morphological sizes and cover rates. In each validation, the architecture was configured as U-Net, employing VGG-11 as the specific backbone, while all other parameters were maintained at their default values as outlined in **Table 2**. **Table 7** and **Table 8** detail the ablation results of various morphological sizes and cover rates. As anticipated, the constrained model consistently exhibited superior performance than the baseline model across varying closing element sizes, dilation element sizes, and coverage rates. Notably, the closing element size of 25×25 and the dilation element size of 20×20 in **Table 7** attained a new state-of-the-art performance, surpassing the default parameter settings [51]. Similarly, the cover rate of 0.90 in **Table 8** presented better performance than the default setting of 0.99. These observations suggest that our proposed constrained training could be further enhanced through hyper-parameter optimization.

**4. Discussion**

Conventional lesion segmentation models typically employed an end-to-end approach, directly mapping input medical images into lesion delineations without accounting for clinical knowledge, such as the spatial distribution of the disease. In this study, we introduced a loss function that incorporates disease occurrence area as a constraint into the standard training framework. Through numerical studies, we demonstrated the effectiveness of the proposed training strategy on pneumothorax segmentation, a condition sensitive to location, manifesting in the pleural space between the lung and the chest wall. Consistently, our proposed training method outperformed the baseline approach across three different architectures paired with two encoders.

While our method was showcased for pneumothorax segmentation, its applicability extends to other thoracic diseases. Many of these conditions exhibit location-sensitive characteristics within the lung area [43, 73, 74]. For thoracopathies localized in the lung area, removing the morphological dilation and utilizing the resulting lung area can potentially improve the model performance using our proposed constrained training strategy. Additionally, the proposed method can be adapted to thoracopathy localization task that require a bounding box around the lesion area pre-specified by expects. A two-step method would be to first generate pixel-level constraints of lung area and identify the rectangular boxes encompassing these constraints, and then to integrate the penalty term in **Formula (3)** into the classic $L2$ distance-based loss function, ensuring the model's output remains within the bounding box-based constraints.

In contrast to the previous constraint learning methods [39, 44, 45, 48, 49] that required additional annotation on the target dataset and task, our method employs a three-phase pipeline to generate sample-specific constraints. This sets a valuable benchmark for the development of constrained models in various domains. Our method can be generalized by (1) generating initial constraints based on an auxiliary task and external datasets, (2) refining these initial constraints based on the relationship between the auxiliary and the target tasks, and (3) employing the target dataset and available annotations to filter out noisy constraints, retaining only the informative ones. As highlighted in **Table 6**, our ablation experiment indicates that the initial constraints transferred from external datasets can introduce detrimental noise, potentially hindering model convergence. However, after refining and selecting constraints, the model's performance improves. **Tables 7** and **8** further demonstrate the robustness of our method to hyper-parameter variations in the second and third stage of learning constraints. In fact, segmenters trained under various settings outperformed the one under the default setting, suggesting a promising avenue for future hyper-parameter optimization.

While the proposed method has demonstrated consistent and robust improvements, there are several limitations to be addressed in future work. First, our model involves multiple hyper-parameters. Although we have made initial adjustments, there is potential for further optimization. Intuitively, we hypothesize that there exists certain relationship between the hyper-parameters [75]. For instance, extensive morphological operations can broaden constraints boundaries, suggesting a need to adjust the cover rate to mitigate noise from these expanded boundaries. Second, while this work primarily incorporates anatomical shape as constraints, future research will delve into other geometric attributes such as sphericity, convexity, and roundness [58, 76, 77]. It is of interest to explore their impact on pneumothorax segmentation. Lastly, previous studies demonstrate a strong correlation between anatomical information and the diagnosis of various thoracic disease [51-54, 78]. Therefore, we plan to evaluate the adaptability of the proposed method across a broader spectrum of thoracopathy tasks, including classification, detection, and segmentation [7, 79-82].

## 5. Conclusion

Historically, domain knowledge was underemphasized by the DL community when tackling medical tasks [83, 84]. In this study, we underscore the value of integrating clinical knowledge, particularly regarding disease occurrence, to enhance DL-based pneumothorax segmentation. Different from previous work that requires additional annotation on the target dataset and task [39, 44, 45, 48, 49], our approach leverages external open-access datasets and an auxiliary task. This strategy not only streamlines our process but also offers versatility, making it a promising framework for diagnosing other thoracic conditions or diseases with location-relevant characteristics [43, 51, 74, 85-87].

**Conflict of interest statement**
All authors declare that they have no conflicts of interest.

**Declaration of generative AI and AI-assisted technologies in the writing process**
During the preparation of the initial draft, HY used GPT-3.5 in order to check grammar. After using this tool, HY and other authors reviewed and edited the content as needed. HY took full responsibility for the content of the publication.

[84] S. Pereira *et al.*, "Enhancing interpretability of automatically extracted machine learning features: application to a RBM-Random Forest system on brain lesion segmentation," *Medical Image Analysis,* vol. 44, pp. 228-244, 2018.

[85] I. Ullah, F. Ali, B. Shah, S. El-Sappagh, T. Abuhmed, and S. H. Park, "A deep learning based dual encoder–decoder framework for anatomical structure segmentation in chest X-ray images," *Scientific Reports,* vol. 13, no. 1, p. 791, 2023.

[86] A. Paul, Y.-X. Tang, T. C. Shen, and R. M. Summers, "Discriminative ensemble learning for few-shot chest x-ray diagnosis," *Medical Image Analysis,* vol. 68, p. 101911, 2021.

[87] A. E. Johnson *et al.*, "MIMIC-CXR, a de-identified publicly available database of chest radiographs with free-text reports," *Scientific Data,* vol. 6, no. 1, p. 317, 2019.

Table 1: An overview of the used dataset, abbreviation, and function

| Dataset name | Abbreviation | Purpose |
|---|---|---|
| Japanese Society of Radiological Technology dataset [46] | JSRT | Lung segmentation |
| Shenzhen dataset [47] | Shenzhen | Lung+ space segmentation |
| Montgomery County dataset [47] | MC | |
| Society for Imaging Informatics in Medicine-American College of Radiology Pneumothorax Segmentation dataset [62] | SIIM-ACR | Lung+ space discrimination |
| | | Pneumothorax segmentation |

Table 2: Default experimental details in constrained segmentation

| Phase | Hyper-parameter | Candidate |
|---|---|---|
| Auxiliary lung segmenter | Architecture | U-Net |
| | Backbone | VGG-11 |
| | Loss function | Dice loss |
| | Optimizer | SGD |
| Lung+ space segmenter | Closing element size | 19×19 |
| | Dilation element size | 15×15 |
| Lung+ space discriminator | Cover rate | 0.99 |
| | Backbone | VGG-11 |
| | Loss function | Cross entropy |
| | Optimizer | SGD |
| Pneumothorax segmenter | Architecture | U-Net, LinkNet, PSPNet |
| | Backbone | VGG-11, MobileOne-S0 |
| | Loss function | Dice loss |
| | Optimizer | SGD |

Table 3: Performance of the lung segmenter on the external lung segmentation test datasets of JSRT, Shenzhen, and MC, in terms of IoU, DSC, and HD. The mean value of each measurement is presented alongside its corresponding standard error, enclosed within brackets. Higher DSC or IoU values indicate better performance, while lower HD values demonstrate better performance.

| Datasets | IoU | DSC | HD |
| --- | --- | --- | --- |
| JSRT | 0.949 (0.003) | 0.974 (0.001) | 3.814 (0.142) |
| Shenzhen | 0.918 (0.008) | 0.956 (0.005) | 4.158 (0.179) |
| MC | 0.961 (0.003) | 0.980 (0.002) | 3.732 (0.153) |

Table 4: Classification results of the lung+ space segmentation discriminator on the test dataset of the SIIM-ACR dataset. The mean value of each metric is presented alongside its corresponding standard error, enclosed within brackets.

| Cutoff value[1] | AUROC | Specificity | Sensitivity | PPV | NPV |
| --- | --- | --- | --- | --- | --- |
| 0.68 | 0.750 (0.023) | 0.854 (0.031) | 0.394 (0.031) | 0.846 (0.030) | 0.410 (0.029) |
| 0.70 | | 0.924 (0.025) | 0.244 (0.026) | 0.867 (0.041) | 0.376 (0.027) |
| 0.70 | | 0.924 (0.025) | 0.244 (0.026) | 0.867 (0.041) | 0.376 (0.027) |
| 0.72 | | 0.962 (0.014) | 0.147 (0.017) | 0.887 (0.038) | 0.358 (0.025) |

[1] Cutoff value was incrementally determined with a step of 0.1, utilizing specificity thresholds of 0.80, 0.85, 0.90, and 0.95 on validation set. In our experiments, as cutoff values were escalated, shifts in specificity values were observed on validation set, resulting in identical cutoff values across distinct specificity thresholds of 0.85 and 0.90.

Table 5: Performance comparison of the constrained and baseline segmentation with different architectures and backbones, in terms of IoU, DSC, and HD. The mean value of each measurement is presented alongside its corresponding standard error, enclosed within brackets. Higher DSC or IoU values indicate better performance, while lower HD values demonstrate better performance.

| Architectures | Backbones | Methods | IoU | DSC | HD |
|---|---|---|---|---|---|
| U-Net | VGG-11 | Baseline | 0.316 (0.010) | 0.441 (0.012) | 4.799 (0.060) |
| | | Ours | 0.336 (0.010) | 0.461 (0.012) | 4.558 (0.053) |
| | | Improvement | 6.3% | 4.5% | 5.0% |
| | MobileOne-S0 | Baseline | 0.309 (0.011) | 0.431 (0.014) | 4.703 (0.050) |
| | | Ours | 0.326 (0.010) | 0.449 (0.011) | 4.586 (0.046) |
| | | Upgrade | 5.5% | 4.2% | 2.5% |
| LinkNet | VGG-11 | Baseline | 0.305 (0.009) | 0.426 (0.011) | 4.740 (0.045) |
| | | Ours | 0.322 (0.011) | 0.447 (0.013) | 4.592 (0.052) |
| | | Improvement | 5.6% | 4.9% | 3.1% |
| | MobileOne-S0 | Baseline | 0.302 (0.009) | 0.425 (0.011) | 4.839 (0.049) |
| | | Ours | 0.320 (0.010) | 0.447 (0.012) | 4.675 (0.044) |
| | | Improvement | 6.0% | 5.2% | 3.4% |
| PSPNet | VGG-11 | Baseline | 0.302 (0.010) | 0.424 (0.012) | 4.866 (0.050) |
| | | Ours | 0.307 (0.009) | 0.429 (0.011) | 4.660 (0.052) |
| | | Improvement | 1.7% | 1.2% | 4.2% |
| | MobileOne-S0 | Baseline | 0.260 (0.009) | 0.377 (0.011) | 5.008 (0.055) |
| | | Ours | 0.267 (0.010) | 0.382 (0.012) | 4.935 (0.053) |
| | | Improvement | 2.7% | 1.3% | 1.5% |
| Average Upgrade | | | 4.6% | 3.6% | 3.3% |

Table 6: Ablation study on the lung+ space segmentation model and the lung+ space segmentation discriminator.

| Methods | lung area segmenter | lung+ space segmenter | lung+ space discriminator | IoU | DSC | HD |
|---|---|---|---|---|---|---|
| Baseline | × | × | × | 0.316 (0.010) | 0.441 (0.012) | 4.799 (0.060) |
| | √ | × | × | 0.298 (0.010) | 0.421 (0.012) | 4.895 (0.053) |
| | × | √ | × | 0.317 (0.010) | 0.439 (0.012) | 4.770 (0.045) |
| Ours | × | √ | √ | 0.336 (0.010) | 0.461 (0.012) | 4.558 (0.053) |

Table 7: Robustness study on morphological element sizes in the lung+ space segmenter.

| Methods | Closing element size | Dilation element size | IoU | DSC | HD |
|---|---|---|---|---|---|
| Baseline | × | × | 0.316 (0.010) | 0.441 (0.012) | 4.799 (0.060) |
| Ours | 15×15 | 10×10 | 0.331 (0.010) | 0.458 (0.012) | 4.691 (0.045) |
| | 15×15 | 15×15 | 0.338 (0.011) | 0.462 (0.012) | 4.439 (0.049) |
| | 15×15 | 20×20 | 0.338 (0.010) | 0.463 (0.012) | 4.538 (0.051) |
| | 19×19[1] | 10×10 | 0.332 (0.010) | 0.455 (0.012) | 4.521 (0.044) |
| | 19×19[1] | 15×15[1] | 0.336 (0.010) | 0.461 (0.012) | 4.558 (0.053) |
| | 19×19[1] | 20×20 | 0.340 (0.010) | 0.465 (0.012) | 4.488 (0.058) |
| | 25×25 | 10×10 | 0.339 (0.011) | 0.463 (0.014) | 4.442 (0.040) |
| | 25×25 | 15×15 | 0.337 (0.009) | 0.465 (0.012) | 4.570 (0.045) |
| | 25×25 | 20×20 | 0.346 (0.011) | 0.473 (0.013) | 4.372 (0.047) |

[1]Parameter settings in the prior literature [51].

Table 8: Robustness study on cover rates in the lung+ space discriminator.

| Methods | Cover rate | IoU | DSC | HD |
|---|---|---|---|---|
| Baseline | × | 0.316 (0.010) | 0.441 (0.012) | 4.799 (0.060) |
| Ours | 0.80 | 0.334 (0.010) | 0.460 (0.012) | 4.408 (0.052) |
| | 0.90 | 0.343 (0.011) | 0.470 (0.013) | 4.474 (0.048) |
| | 0.99 | 0.336 (0.010) | 0.461 (0.012) | 4.558 (0.053) |

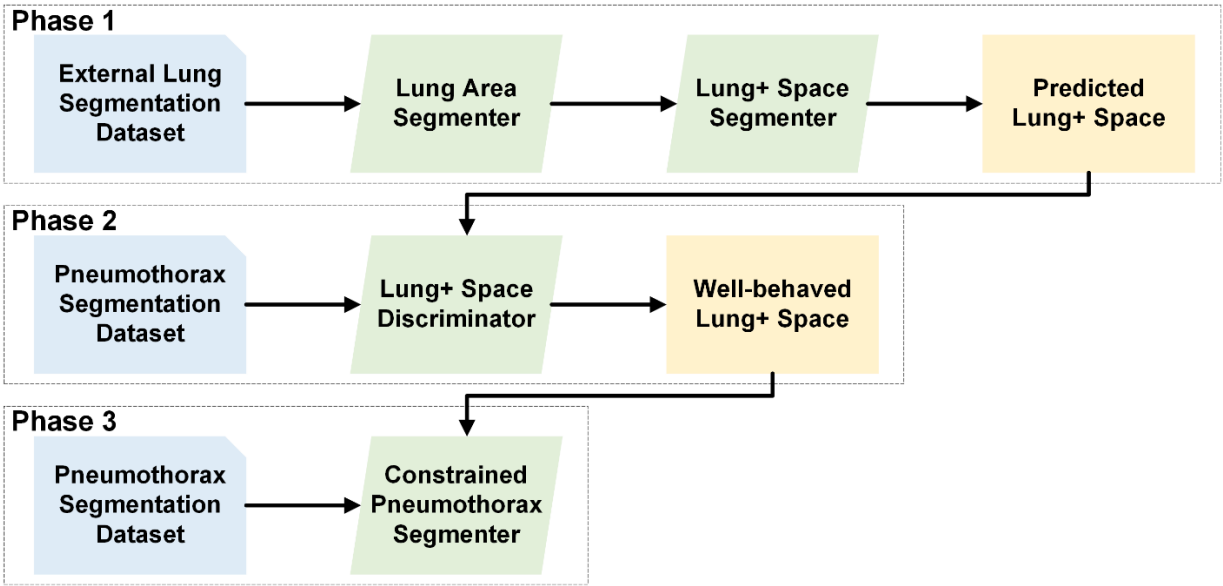

Figure 1: Schematic diagram of the proposed segmentation framework.

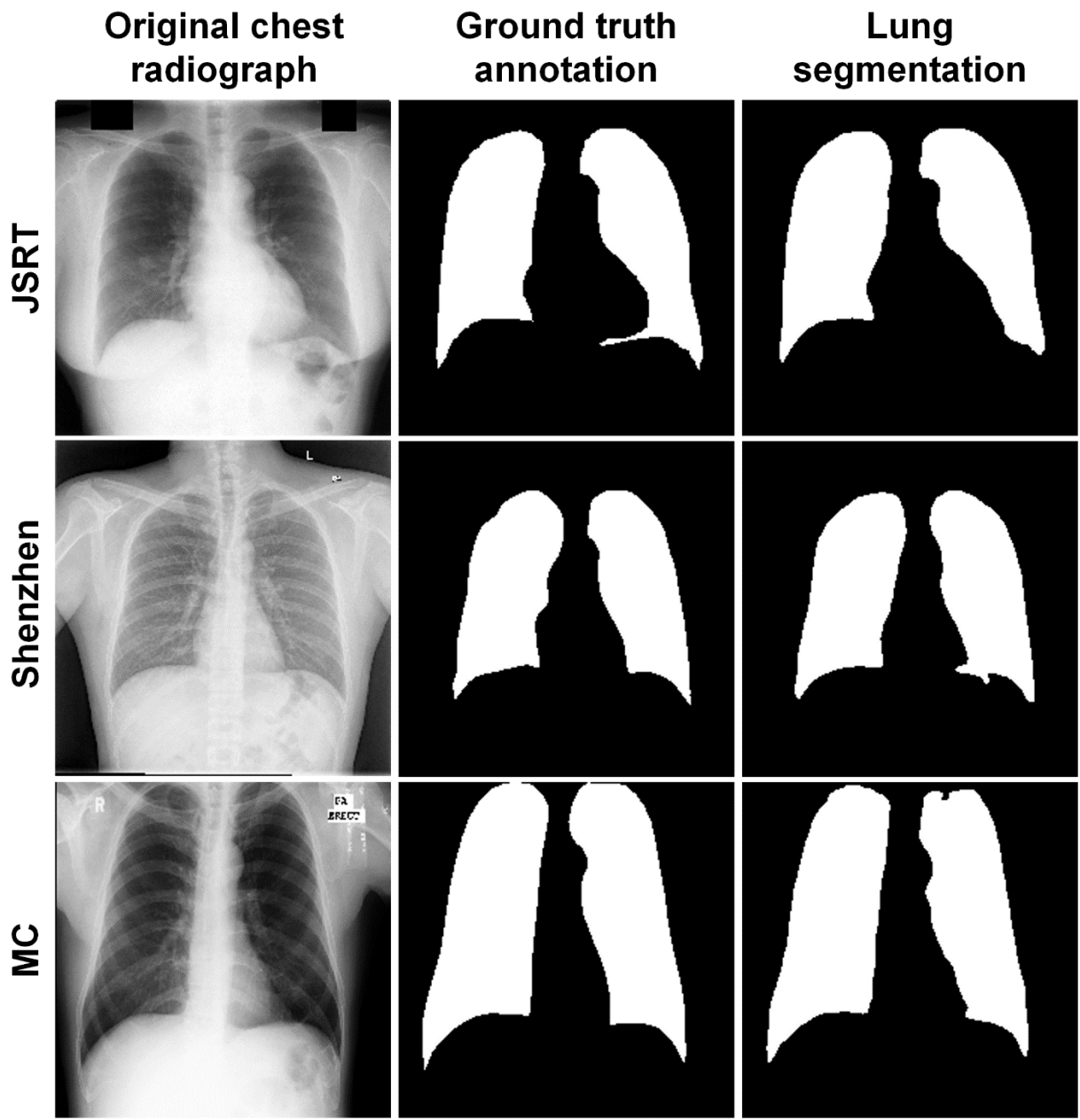

Figure 2: Visual samples of lung area segmentation in JSRT, Shenzhen, and MC databases.

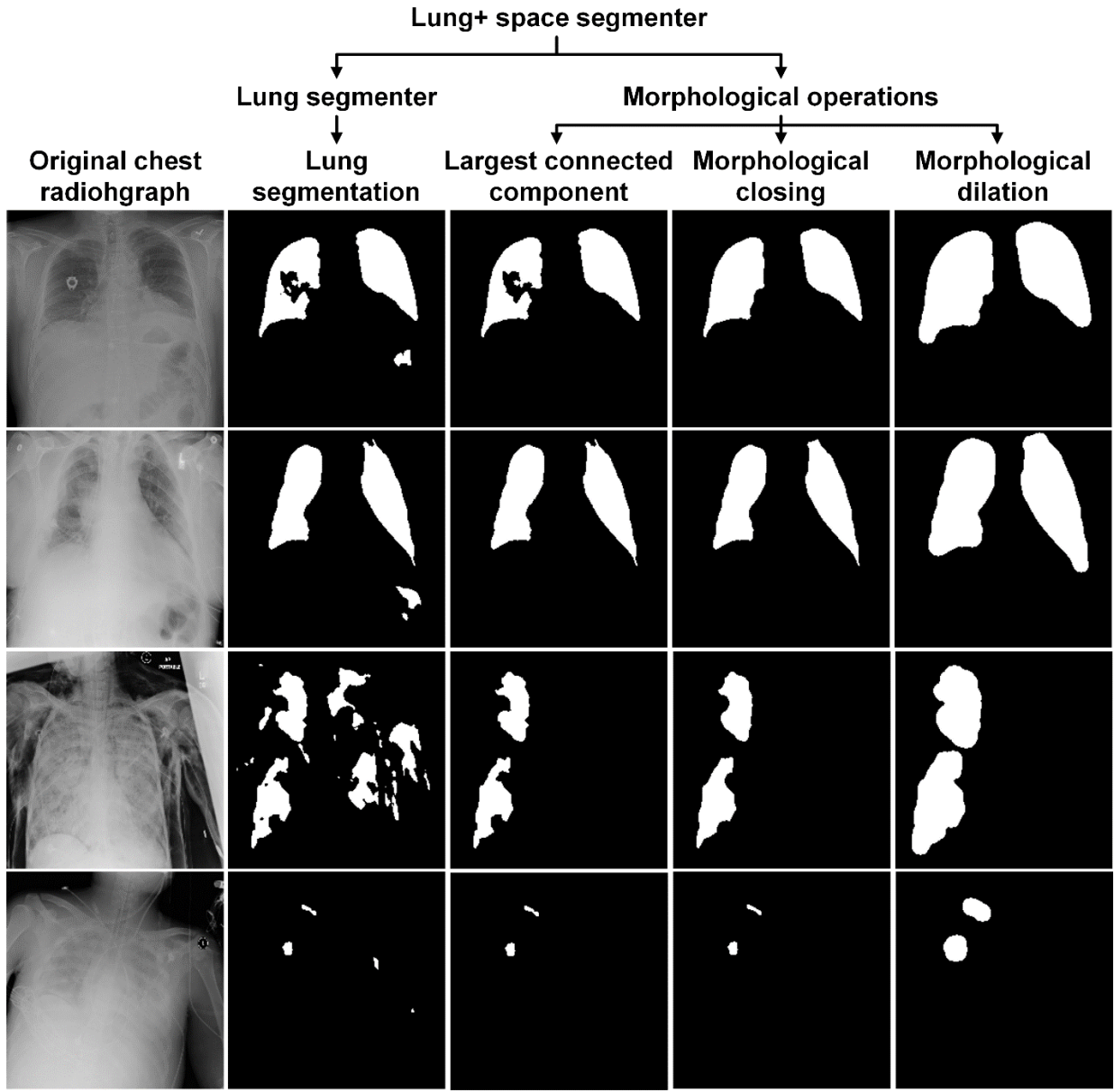

Figure 3: Visualization of different generation stages of the constraints of lung+ space in SIIM-ACR database.

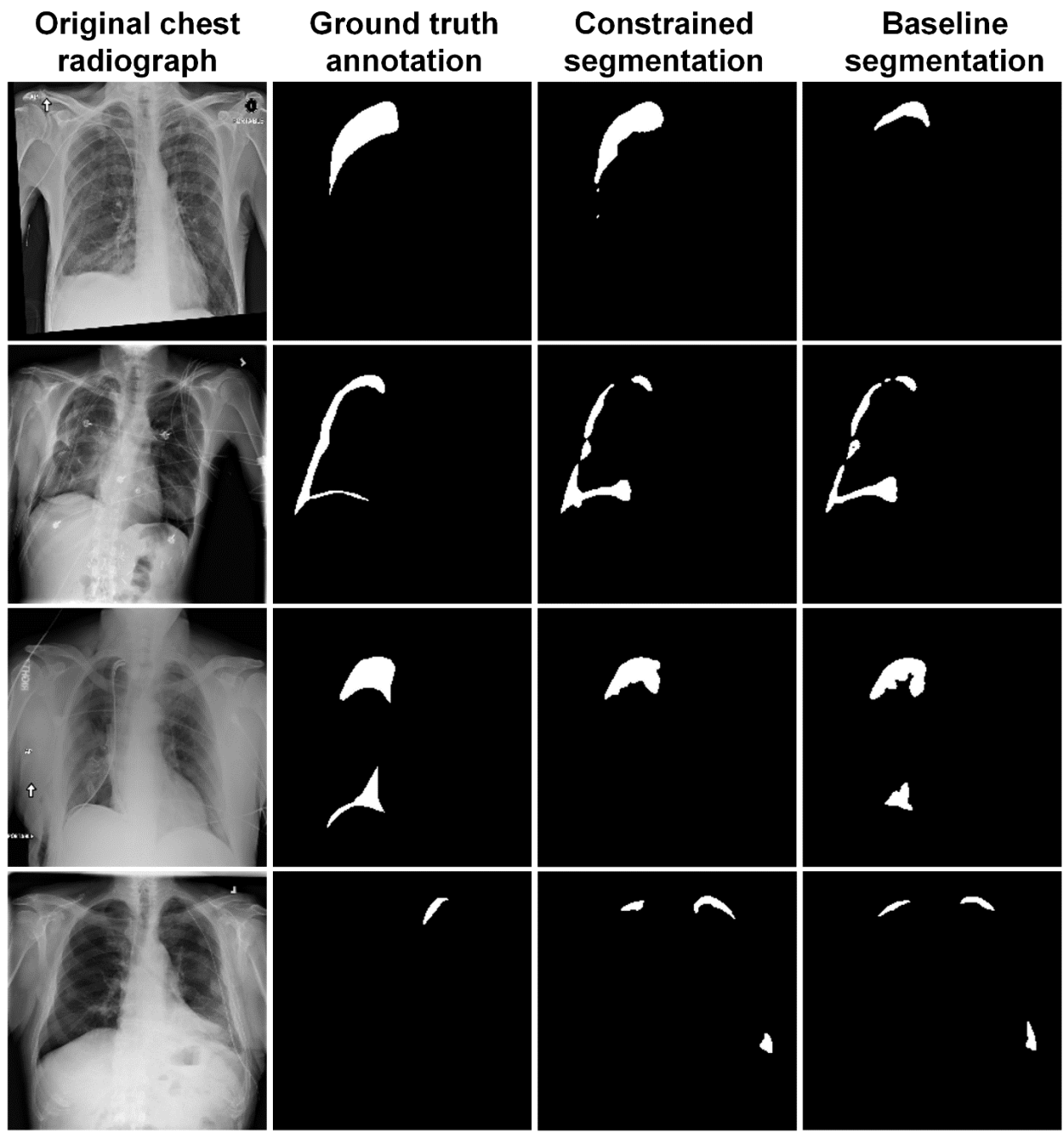

Figure 4: Comparative example of the constrained segmentation and the baseline segmentation. Rows 1 to 4 represent the following scenarios: the constrained method outperforming the baseline, the constrained method achieving comparable results to the baseline, the constrained method underperforming the baseline, and both methods collapsing, respectively.